\documentclass[a4paper,11pt]{article}
\pdfoutput=1 % if your are submitting a pdflatex (i.e. if you have
\usepackage{jheppub} % for details on the use of the package, please
\usepackage[T1]{fontenc} % if needed
\usepackage[dvipsnames]{xcolor} %Must come before tikz!
\usepackage[stable]{footmisc}
\colorlet{cite}{LimeGreen!50!Green}

\title{\boldmath Higher form symmetries, membranes and flux quantization}

  \author[a]{F. Caro P\'erez,} %\footnote{The order of the  authors is alphabetical.}}
 \author[b]{M.P. García del Moral,}
 \author[a]{A. Restuccia,} 

\affiliation[a]{Universidad de Antofagasta,\\Departamento de Física, Universidad de Antofagasta, Aptdo 02800, Chile}
\affiliation[b]{Universidad de la Rioja,\\Área de Física, Departamento de Química, Centro Científico Tecnológico Universidad de la Rioja, La Rioja, 26006, Spain}
 
\emailAdd{fabian.caro.perez@ua.cl}
\emailAdd{m-pilar.garciam@unirioja.es}
\emailAdd{alvaro.restuccia@uantof.cl}

\abstract{Higher Forms Symmetries (HFS) of a closed bosonic M2-brane theory formulated on a compactified target space $\mathcal{M}_9 \times T^2$ are obtained. We show that the cancellation of the 't Hooft anomaly present in the theory is related to a 3-form flux with $\mathcal{G}_1^{\nabla}$-gerbe structure associated to the world-volume flux quantization condition. A Wilson surface  is naturally introduced  on the topological operator that characterize the holonomy of the M2-brane. The projection of the flux quantization condition inherited from the gerbe structure onto the spatial part of the worldvolume, leads to a flux quantization on the M2-brane. The topological operators realise discrete symmetries associated with the winding and the flux/monopole condition. The algebra of operators is well defined.}

\begin{document}
%\emergencystretch 3em
%\hypersetup{pageanchor=false}
%\makeatletter
%\let\old@fpheader\@fpheader
%\preprint{IFT-UAM/CSIC-24-68}
\makeatother
\maketitle
\flushbottom
\noindent
\section{Introduction} 
In this paper we study the relation between Higher form symmetries, the anomaly cancellation and the quantization conditions over the bosonic M2-brane. We show that the cancellation of the t'Hooft anomaly associated with the existence of a Higher Form Symmetry (HFS) on the bosonic M2-brane implies the existence of a flux condition over its worldvolume. This result is a necessary condition for the consistency of the theory. It is noteworthy that  the spectral analysis of the M2-brane compactified on toroidal backgrounds also requires the imposition of a flux condition  for the discreteness of the mass spectrum of the toroidally compactified supermembrane, 
\cite{GarciaDelMoral:2018jye}, \cite{GarciadelMoral:2020dfs}. 
\newline
Symmetries are fundamental in physics. They have allowed to unify forces but also to characterize phases of matter. They can be defined as transformations under which, the action describing a physical system, remains invariant. These transformations are characterized by groups. These groups can be either finite or infinite dimensional. Associated to the continuous symmetries, there are conserved quantities and the generators of the symmetry are described by local operators. In this framework the Noether symmetries can be denoted as 0-symmetries.
A limitation of Noether symmetries is that they cannot indicate the presence of a spontaneous symmetry breaking or a phase transition.
\newline
Generalized symmetries represent a vast concept that generalises the notion of symmetry. In this paper, the focus will be on Higher Form symmetries of a given order, for example $p$, denoted in the literature as $HFS(p)$. These symmetries are understood to act on non-local operators and are associated with the existence of higher-order charges. The operators are denoted as invertible operators when they are topological and gauge-invariants; otherwise, they are referred to as non-invertible operators.\newline
The HFS are useful to classify topological phases in condensed matter physics \cite{Armas:2023tyx} and in the context of High Energy Physics.  They arise from the cancellation of a new type of anomalies that involve non-dynamical fields.These fields, denoted as background fields, are topological in the target manifold but can be dynamical in manifolds of higher dimensions. \newline
 Recently, there has been a growing interest in exploring the potential applications of Higher Form Symmetries to various gauge and gravitational theories.
See for example in the context of supergravity {\cite{Zhang:2023wai}, string theory \cite{Bhardwaj:2023kri}, \cite{Heckman:2024obe}, F-theory \cite{Cvetic:2021sxm} and M-theory \cite{Albertini:2020mdx} among others. The existence of these new Higher Order Symmetries has been linked to the physical consistency of a theory. Their presence implies the disappearance of several anomalies. To illustrate this point, consider the Maxwell case, where the presence of these symmetries leads to the cancelation of the 't Hooft anomaly \cite{Gaiotto:2014kfa}. A similar phenomenon occurs in supergravity, where the cancelation of the ABJ anomaly \cite{Garcia-Valdecasas:2023mis} is implied. The topological operators of the theory define new topological charges and the original gauge symmetry groups in the presence of these charges break generically into their discrete subgroups. Interestingly, this is an expected property for well-defined quantum gravity theories. 
\newline
The objective of this paper is to demonstrate the existence of HFS associated with the bosonic sector of the M2-brane. The partition function associated with this HFS exhibits a 't Hooft anomaly. This anomaly is cancelled through the introduction of background fields with support on a 4D manifold, which couple to the membrane 3D action via a BF coupling and inflow terms. This anomaly cancellation gives rise to a geometrical structure of a $\mathcal{G}_1^{\nabla}$-gerbe on the M2-brane theory (see Appendix \ref{APENDICE GERBE}} for details of the notation of gerbe structure). The projection of this structure through the worldvolume gives rise to a flux term on the membrane. The flux quantisation condition guarantees the quantisation of the charges and is associated with the existence of a monopole over the membrane. It is noteworthy that the supermembrane with worldvolume fluxes, i.e.  monopoles \cite{Martin:1997cb}, has a purely discrete supersymmetric spectrum with finite multiplicity \cite{Boulton:2002br}, \cite{Boulton:2010nd}. This property also holds in the presence of quantized three-forms whose pullback induces, the worldvolume flux conditions over the M2-brane \cite{GarciaDelMoral:2018jye}, \cite{GarciadelMoral:2020dfs}.
\newline
It has been established since the early nineties that bosonic p-branes with $p\ge 2$ are quantum inconsistent for any dimension, in clear contrast to the case of the bosonic string, which is consistent for $d=26$. The theory contains Lorentz anomalies due to order ambiguities at the quantum level, which cannot be avoided in any gauge, not even in the Light Cone Gauge (LCG) \cite{Bars:1987nr}, \cite{Bars:1989ba}. This problem is avoided by the introduction of supersymmetry.  The supersymmetric M2-brane can be formulated consistently on certain dimensions, \textit{i.e.} in d=4,5,7 and 11 dimensions as demonstrated in \cite{Duff:1987cs}. At the supersymmetric level, the action can be made Lorentz anomaly free for $D=11$, by considering its formulation on a Light Cone Gauge (LCG) and imposing kappa symmetry, \cite{Marquard:1989rd}. The resulting theory, formulated a la Green-Schwarz, is invariant under area preserving diffeomorphisms. This is the  residual symmetry of the M2-brane in the LCG formulation. 
\newline
It is known that the supermembrane, when formulated on an 11D Minkowski spacetime, has a continuum spectrum from $(0, \infty)$ \cite{deWit:1988xki}. This behaviour is general for the Supermembrane theory on arbitrary backgrounds and it does not change  simply by compactification, for example on a torus. Classically, this is instability is understood as the presence of string-like spikes with zero energy cost, which avoid the preservation of the topology nor the number of membranes, an aspect that leads to interpret the theory as a theory of interactions of D0-branes, a second quantized theory. This property of the supermembrane mass spectrum holds for most of the backgrounds. However, there are also backgrounds where this is not the case and the spectrum is purely discrete, with finite multiplicity \cite{Boulton:2002br} \cite{Boulton:2010nd} . In such cases, the theory can be regarded as a first quantized theory, capable of capturing some of the microstates of M-theory. It is reasonable to consider that these behaviours may emerge as a consequence of the cancellation of an as yet unidentified anomaly of the M2 membrane. This possibility has not been considered in the past. Our analysis indicates that this is indeed the case. In this paper we identify the existence of a 't Hooft anomaly in bosonic theory, whose cancellation requires the existence of a HFS inducing a $\mathcal{G}_1^{\nabla}$-gerbe over the M2-brane and generating new discrete symmetries associated with the presence of a world-volume flux on the membrane. We find that this implies a flux quantization condition needed to renders the spectrum of the supersymmetric M2-brane compactified on toroidal backgrounds, discrete. 
\newline
The paper is structured as follows: in Section 2 we briefly review Higher Form Symmetries, taking as an example the case of Maxwell theory in $\mathbb{R}^2\times T^2$. In Sec. 3 we present the new results: the HFS of the bosonic M2-brane theory which cancels the 't Hooft anomaly in 11D. In section 4 the algebra of operators is found. Section 5 ends with a brief discussion and conclusions. In appendix \ref{APENDICE DEMOSTRACION FREE ANOMLA TERM}, the anomaly cancellation of the M2-brane path integral in the presence of background fields is shown. In appendix \ref{APENDICE GERBE}, we introduce the structure of $\mathcal{G}_1^{\nabla}$-Gerbe that is used in section 3.

\section{Higher form symmetries for gauge theories}\label{Capitulo HFS}
In this section, we briefly introduce Higher Form Symmetries (\textit{HFS}$(p)$) and perform the calculation for the abelian $U(1)$ gauge theory on $\mathbb{R}^2 \times T^2$. 
In a standard QFT, the Noether symmetries of the theory are associated with unitary generators acting on codimension-1 objects. To generalize this, one considers non-local topological operators acting on codimension-$(p+1)$ objects, associated to the symmetry group $G$. We denote them by $U_g$ where the labell $g \in G$ \cite{Bhardwaj:2023kri}, \cite{Gaiotto:2014kfa}, \cite{Gomes:2023ahz}, \cite{Schafer-Nameki:2023jdn}. These operators are topological if they remain invariant under smooth deformations of the integration submanifold, as follows:
\begin{align}\label{OPERADORES GENERALIZADOS}
    U_{g}(\Sigma_{d-p-1}) = U_{g}(\eta(\Sigma_{d-p-1})),
\end{align}
where $\eta: \Sigma_{d-p-1} \to \Sigma_{d-p-1}'$ is a smooth deformation of $\Sigma_{d-p-1}$ such that:
\begin{align}
    \chi(\eta \circ\Sigma_{d-p-1}) = \chi(\Sigma_{d-p-1}).
\end{align}
 With $\chi(\circ)$ representing the Euler-Characteristic of the (sub)manifold considered. This is only possible if there exist a $(d-p)-$manifold $\mathcal{Y}_{d-p} \in \Omega_{0}^G$
 \footnote{The \textit{cobordism group} $\Omega_q^G$ is a graded group that classifies manifolds up to the cobordism relation. Two closed $(d-p-1)$-dimensional manifolds $\Sigma_{d-p-1}$ and $\Sigma'_{d-p-1}$ are said to be \textit{cobordant} if there exists a compact $(d-p)$-dimensional manifold $\mathcal{Y}_{d-p}$ such that $\partial \mathcal{Y}_{d-p} = \Sigma_{d-p-1} \sqcup (\Sigma'_{d-p-1})$. The set of all such equivalence classes forms the cobordism group $\Omega_q^G$, where the operation is given by disjoint union.}

\begin{align}
    \partial \mathcal{Y}_{d-p} = \Sigma_{d-p-1} \sqcup \Sigma_{d-p-1}',
\end{align}
where \( \Sigma_{d-p-1}' \) has the opposite orientation with respect to $\Sigma_{d-p-1}$. Furthermore, the operators in \eqref{OPERADORES GENERALIZADOS} act on each other according to fusion rules inherited from the gauge group of the field theory formulated on the base manifold $\mathcal{M}$. Let $g, g'\hspace{1mm} \& \hspace{1mm} gg' \in G$, then we have:
\begin{align}\label{PROPIEDAD INVERTIBILIDAD}
    U_{g}(\Sigma_{d-p-1}) U_{g'}(\Sigma_{d-p-1}) = U_{gg'}(\Sigma_{d-p-1}), 
\end{align}
where $U_g \in \mathbb{G}^{(p)}$. $\mathbb{G}^{(p)}$ is the global symmetry group associated to the HFS(p), being $p$ the rank of the HFS where the topological operators act. \newline
In a QFT, we have two possible cases for \textit{HFS}: invertible and non-invertible symmetries. A generalized symmetry is invertible if the topological operators are gauge invariant. 
The second case occurs when the symmetry generators are either  not topological or either not gauge invariant and therefore do not satisfy the property \eqref{PROPIEDAD INVERTIBILIDAD}. Their analysis is different but will not be addressed in this work. For a discussion see,\cite{Garcia-Valdecasas:2023mis}, \cite{Karasik:2022kkq}, \cite{Choi:2022jqy}, \cite{Fernandez-Melgarejo:2024ffg}, \cite{Hasan:2024aow}. \newline
The invertible \textit{HFS}$(p)$ are obtained by calculating the generalization of the Noether current 
$\star J_{d-p-1} \in \Omega^{p+1}(\mathcal{M})$, whose associated form is closed in the \textit{De Rham} cohomology sense, this is equivalent to require that the current follows a conservative law structure \textit{á la} Noether:
\begin{align*}
   \star J_{d-p-1} \in Z^{p+1}(\mathcal{M},\mathbb{R}).
\end{align*}
Using the surjective map $G \to Alg(G)=\mathfrak{g}$ with $g=e^{i\alpha}$, we can define the invertible topological operator as follows \cite{Gaiotto:2014kfa}:
\begin{align}\label{OPERADORES U}
      U_{\alpha}(\Sigma_{p+1}) \triangleq {}& \exp \big(i\alpha \int_{\Sigma_{p+1}} \star J_{d-p-1}  \big).
\end{align}
 To visualize this procedure, let us consider a typical example. The case of a free Maxwell theory; denoted by:
\begin{align}
    \big[ U(1), \mathfrak{u}(1), 4, \text{sig}(0,4) \big]-~&\text{Maxwell theory}.
\end{align}
Where we have followed the standard notation in HFS theories \cite{Gaiotto:2014kfa}:
\begin{align}
    \big[ G, \mathfrak{g}, \dim(\mathcal{M}), \text{sig}(q, \dim(\mathcal{M}) - q) \big],
\end{align}
$G$ corresponds to the gauge group of the theory, $Alg(G) = \mathfrak{g}$ denotes its  Lie algebra, then follows the dimension of the base manifold, and its signature. 
In this case, the standard symmetries of  Maxwell theory, \textit{HFS}$(0)$  and  \textit{HFS}$(1)$ that generate a mixed 't Hooft anomaly in the path integral between the electric-magnetic symmetry.  This anomaly is reflected in the fact that is not possible to gauge the global symmetry present in the electric and magnetic sectors of the theory simultaneously.
 We will review it in detail, below, since it is useful for the object of our study in this paper, namely the bosonic part of the M2-brane.
 \newline
 \newline
Let us consider for simplicity, a Riemaniann manifold $(\mathcal{M},\eta)$ with $dim( \mathcal{M})=4$ and a 1-form connection $A \in \Omega^{1}(\mathcal{M})\otimes \mathfrak{u}(1)$ whose associated field strength is $F=dA \in  \Omega^2(\mathcal{M})\otimes \mathfrak{u}(1)$. The Maxwell action without any coupling to matter, is given by:
\begin{align}\label{ACCION DE MAXWELL}
    S_{M}(A)=\frac{1}{2g^2}\int_{\mathcal{M}}F\wedge \star F.
\end{align}
Being $g^2$ the electromagnetic coupling constant, \eqref{ACCION DE MAXWELL} describes an Abelian gauge theory with gauge group $G  \triangleq U(1) $ and a global symmetry $\mathbb{G}^{(1)}$ associated with a flat connection shift. 
The equation of motion of this action is obtained from the variations of the action with respect to dynamical field $A$:
\begin{align}\label{EQUACIONES DE MOVIMIENTO MAXWELL}
    d\star F=0.
\end{align}
Additionaly, the \textit{Bianchi} identity is:
\begin{align}
    dF=0.
\end{align}
Following the procedure for \textit{HFS}$(p)$, the $2-$forms currents are given by $J_e=\frac{1}{g^2}F$ and $J_m=\frac{1}{2\pi}\star F$, where the proportionality constants in the definition of the currents are the conventional ones. From here we can see that $(\star J_e,\star J_m) \in Z^{2}(\mathcal{M},\mathbb{R})$. We can construct the generators of the $1$-form symmetry for this theory by following the structure given by \eqref{OPERADORES U}:
\begin{align}\label{OPERDAROES DE MAXWELL}
    U_\alpha^{(e)}(\Sigma_2)=\exp\big(\frac{i\alpha}{g^2} \int_{\Sigma_2}\star F \big), \quad
    U_\beta^{(m)}(\Tilde{\Sigma}_2)=\exp \big( \frac{i\beta}{2\pi}\int_{\Tilde{\Sigma}_2} F \big).
\end{align}
Where \( \Sigma_2 \) and \( \tilde{\Sigma}_2 \) are 2-dimensional submanifolds of \( \mathcal{M} \).
These operators are gauge-invariant and $U_{\alpha}^{(e)}$ is topological on-shell, while $U_{\beta}^{(m)}$ is topological off-shell. Let \( \eta \) be a smooth deformation homeomorphism:
\begin{align}
    \eta: \Sigma_2 \longrightarrow \Sigma_2',
\end{align}
thus, we can define a non-trivial element \( \mathcal{Y}_3 \in \Omega_{0}^G \) such that \( \partial \mathcal{Y}_3 = \Sigma_2' \sqcup \Sigma_2 \). We assume that the equation of motion \eqref{EQUACIONES DE MOVIMIENTO MAXWELL} can by extended to \( \text{supp}(\mathcal{Y}_3)\). The electric operator transforms as follows: 
\begin{align}
    \eta \circ U_{\alpha}^{(e)}(\Sigma_2) ={}& \exp \big( i\alpha \int_{\Sigma_2^{'}=\eta(\Sigma_2)} \star F \big) = \exp \big( i\alpha \int_{\partial \mathcal{Y}_3 + \Sigma_2} \star F \big) \\ \notag
    ={}& \exp \big( i\alpha \int_{\mathcal{Y}_3} d\star F \big) U_{\alpha}^{(e)}(\Sigma_2) = U_{\alpha}^{(e)}(\Sigma_2).
\end{align}
Similarly, for the magnetic operator. Consequently, the global symmetry group of this theory is given by:
\begin{align}
    \mathbb{G}^{(1)}=U(1)^{(1)}_e  \oplus U(1)_m^{(1)}.
\end{align}

%%%%%%%%%%%%%%%%%%%%%%%%%%%%%%%%%%%%%%%%%%%%%%%%%%%%%%%%%%%%%%%%%%%%%%%%%%%%%%%
\subsection{Anomaly cancellation in Maxwell Theory}
Now, we can perform a gauging of the global symmetry $\mathbb{G}^{(1)}$ by introducing background fields in $\mathcal{M}$. Consider two background fields $(B,\Tilde{B})$ in the cohomology group $H^{2}(\mathcal{M}, \mathbb{R})\otimes H^2(\mathcal{M},\mathbb{R})$ such that, the gauge class of these fields is given by:
\begin{align}
    B \to B+d\Lambda, \qquad \Tilde{B} \to \Tilde{ B} +d\lambda.
\end{align}
In order to gauge the global symmetry, we impose $A \to A+\Lambda$ and $B \to B +d\Lambda$ with the same parameter $\Lambda$. The background fields $B$ and $\tilde{B}$ can be understood as 2-form gauge flat connections, in a broader geometric structure that generalizes the concept of fiber bundles to higher-order forms, known as a Gerbe. The cohomology that classifies these mathematical structures is the Deligne cohomology.  
\newline
In Appendix \ref{APENDICE GERBE} , we provide a summary of the definition and the main properties of Gerbes that will be used in this paper. Consequently, $B$ and $\widetilde{B}$ act as connections of a trivial Gerbe and generate a specific class within Deligne cohomology \cite{Murray:1994db}. 
The coupling in the action is given by:
\begin{align}\label{ACCION DE MAXWELL CON CAMPOS DE BACKGROUND}
    S_{M}(A, B, \Tilde{B})= \frac{1}{2g^2} \int_{\mathcal{M}}\big(F-B \big)\wedge \star \big(F-B \big) +\frac{1}{2\pi}\int_{\mathcal{M}} \Tilde{B}\wedge F.
\end{align}
However, this theory  suffers of a mixed 't Hooft anomaly, which can be visualized in the path integral \(\mathcal{Z}(A, \{B_{I}\})\), where $\{B_{I}\}$ is the set of background fields associated with the $I$-th \textit{HFS}$(p)$ of the theory \cite{Schafer-Nameki:2023jdn}:
\begin{align}
    \mathcal{Z}([A], \{[ B _{I} ]\}) = \phi(\Lambda, \{ B_{I}\}) \times \mathcal{Z}(A, \{B_{I}\}),
\end{align}
where \(\phi(\Lambda, \{B_{I}\})\) is a \(U(1)\)-phase. To eliminate this phase, it is necessary to include an \textit{anomaly theory} contained in \(\mathcal{D}\) with $\dim(\mathcal{D})=5$ and $\partial\mathcal{D}=\mathcal{M}$, such that a new path integral is defined as follows:
\begin{align}
    \tilde{\mathcal{Z}}(A, \{B_{I}\}) = \mathcal{Z}(A,\{B_{I}\}) \times e^{\mathcal{T}_{TQFT}}.
\end{align}
In this way, \(\mathcal{T}_{TQFT}\) \cite{Restuccia:1998yx} is an action that depends only on the background fields with non-trivial \(supp(\mathcal{D})\) and transforms as $\phi(\Lambda, \{B_{I}\})^{-1} \times e^{\mathcal{T}_{TQFT}}$. Hence, an anomalous \textit{inflow} term with support in $\mathcal{D}$ is introduced in the action to cancel the previous anomalous term and to generate a free-anomaly action. The BF-term is given by, 
\begin{align}
  \mathcal{T}_{TQFT} {}& \triangleq  S_{inflow} = -\frac{i}{2\pi}\int_{\mathcal{D}} B \wedge d\Tilde{B}.
\end{align}
 The complete action \eqref{ACCION DE MAXWELL CON CAMPOS DE BACKGROUND} 
 with the new inflow contribution generate a partition function which is anomaly free. The action is invariant under the equivalence class of \([A], [B]\), and \([\Tilde{B}]\).  Therefore, the final free 't Hooft anomaly action is given by:
\begin{align}
    S_T(A,B,\Tilde{B})={}&\frac{1}{2g^2}\int_{\mathcal{M}}\big(F-B \big)\wedge \star \big(F-B \big)+\frac{1}{2\pi}\int_{\mathcal{M}} \Tilde{B}\wedge F- \\ \notag {}& -\frac{1}{2\pi}\int_{\mathcal{D}}B \wedge d\Tilde{B}.
\end{align}
For the anomaly-free action it is possible to define the symmetry operators in the presence of background fields. Observe, that only the magnetic operator receive modifications. 
The global symmetry group after the introduction of the background fields remains preserved, i.e.
\begin{align}
    \mathbb{G}^{(1)} \cong U(1)^{(1)}_e  \oplus U(1)_m^{(1)}.
\end{align}
However, Maxwell theory as a complete theory requires the introduction of matter through point-like charges located at specific points in space. These charges generate singularities. The space where these charges act, is given by the total spacetime minus a point where the charge is located. 
We take the previusly introduced submanifolds $\Sigma_2$ and $\tilde{\Sigma}_2$ as $\mathbb{S}^2$ and $\tilde{\mathbb{S}}^2$ respectively for the electric and magnetic operators. The associated topological operators, in the presence of charges, become:
\begin{align}
     U_\alpha^{(e)}(\mathbb{S}^2)={}& \exp\big(\frac{i\alpha}{g^2} \int_{\mathbb{S}^2}\star( F -B\big))=\exp\big(\frac{i\alpha}{g^2} n), \\
    U_\beta^{(m)}(\Tilde{\mathbb{S}}^2)= {}& \exp \big( \frac{i\beta}{2\pi}\int_{\tilde{\mathbb{S}^2}} F \big)=\exp \big( \frac{i\beta}{2\pi}m),
\end{align}
 with $n,m\in \mathbb{Z}$.
The quantum consistency of the theory demands charge quantization, which imposes a flux condition associated with a non-trivial U(1) gauge bundle over a sphere, with its center at the charge that generates the singularity.
Following the argument of \cite{Hull:2024bcl} and \cite{Luo:2023ive}, the charge of the magnetic sector in the theory becomes ill-defined due to the breaking of the cohomology group of $[B]$ to $\mathbb{Z}$. This occurs because $dB \neq 0$ for a generic background field. Consequently, the magnetic operator in \eqref{OPERDAROES DE MAXWELL} is no longer topological, leading to the breaking of the magnetic symmetry in the presence of background electric fields in $H^{2}(\mathcal{M},\mathbb{R})$. 
It is straightforward to see this via the redefinition of the magnetic charge in the presence of the background field as follows:
\begin{align}
    Q_{m}(A) \to Q_{m}(A,B) =\frac{1}{2\pi} \int_{\tilde{\Sigma_2}}(F-B).
\end{align}
Via the Weil theorem \cite{Brylinski:1993ab}, if we consider $\tilde{\Sigma}_2$ to be a compact oriented surface and \( F \) as the curvature of a connection, then this integral yields \(\mathbb{Z}\), but not necessarily \(\int_{\tilde{\Sigma}_2} B\). This is because \( B \) is an element in the cohomology \( H^{2}(\mathcal{M},\mathbb{R}) \).
Hence this forces the background $B$ to be a quantized closed form to be consistently coupled in the presence of matter.
The global symmetry in this case correspond to 
\begin{align}
    \mathbb{G}^{(1)}=U(1)^{(1)}_e\oplus U(1)^{(1)}_m \to \mathbb{Z}_{N,(e)} \oplus \mathbb{Z}_{M,(m)}.
\end{align}
We remark that this symmetry breaking only happens in the Maxwell theory formulated on a non compact target space when the theory is coupled to matter.

\subsection{\texorpdfstring{4D Abelian gauge theory on a compact spacetime: $\mathbb{R}^2 \times T^2$}{4D Abelian gauge theory on a compact spacetime: R squared x T squared}}
Consider a product base manifold as follows. We add two non-trivial cocycles and decompose the 4D manifold as $2D + 2D$, associated with the non-compact and compact sectors, respectively, under the assumption of Euclidean signature. Due to the Cartesian structure of the product, the action of $U(1)-$gauge theory separates additively in the base manifold only if $\tilde{A} \in \Omega^1(T^2)$ is a function of $x^r$ and $A \in \Omega^1(\mathbb{R}^2)$ is a function only of $x^i$:
\begin{align}
    S_{M} \to S_{M}(\tilde{A},A)= {}& \frac{1}{2g^2}\int_{T^2}\tilde{F}\wedge \star_{T^2} \tilde{F} + \frac{1}{2g^2}\int_{\mathbb{R}^2}F\wedge \star_{\mathbb{R}^2} F.
\end{align}
Here, the indices are $r, s,\in T^2$ and $i, j  \in \mathbb{R}^2$.
The equations of motion give the standard Maxwell equations, but take separate forms for the non-compact and compact sector of the spacetime. For the non-compact sector of the action, the equation of motion and the \textit{bianchi} identity are given by:
\begin{align}
    d\star_{\mathbb{R}^2} F={}& 0,\\
    dF={}&0,
\end{align}
and for compact sector by:
\begin{align}\label{EQ DE MAXWELL EN T^2}
    d\star_{T^2} \tilde{F}={}&0,\\
    d\tilde{F}={}&0. \label{EQ 2.31}
\end{align}
The second equation \eqref{EQ 2.31} together with $\int_{T^2} F \in 2\pi \mathbb{Z}$, via the Weil theorem \cite{Brylinski:1993ab}, enssures the existence of a $P[U(1)]-$bundle\footnote{In our notation, a $P[G]-$ bundle is a principal bundle with structure group $G$.} over $T^2$, that permits the classification of fiber bundles within the cohomology group $H^2(T^2, \mathbb{Z}) \cong \mathbb{Z}$ \cite{Brylinski:1993ab}. 
For the non-compact sector of the theory, the generators of symmetry $U_{\alpha}(\Sigma_{3-p})$ are trivial because $\mathbb{R}^2$ is contractible to a point.
The topological operators of the compact sector are given by the following expressions:
\begin{align}
    U_{\alpha}^{(e)}(\Sigma_0 \sim \{ q \})={}& \exp (2\alpha K(T^2)), \\
    U_{\beta}^{(m)}(T^2)={}& \exp \bigg(i\beta \int_{T^2} F \bigg),
\end{align}
where $\{ q\}$ is a generic point, and $K$ is a map: $\Sigma_0 \subset T^2 \to \mathbb{Z}$. The same arguments as in the preceding section can be used to include background fields which cancel the 't Hooft anomaly. However, due to a volume factor, arising from the non-compact sector, that factorizes in the path integral, the computation must be performed in a box.
\section{Higher form symmetries  of the bosonic M2-brane}

The action of a bosonic $\tilde{p}$-brane\footnote{There is no correlation between the   $\tilde{p}$ and the $p$ appearing in \textit{HFS}$(p)$.}, \textit{á la} Nambu Goto \cite{Goto:1971ce}, \cite{Nambu70}, formulated on a flat spacetime is given by: 
\begin{equation}
S_{NG}(X)=-T_{M2}\int_{\Sigma_{\tilde{p}+1}} d^{{\tilde{p}}+1}\xi[ \det(\partial_i X^{M}\partial_jX_{M})]^{\frac{1}{2}},
\end{equation}
where the dynamic fields correspond to the embedding maps from the $\tilde{p}$-brane worldvolume  $\Sigma_{\tilde{p}+1}$ onto the target space: $X^{M}:\Sigma_{\tilde{p}+1}\to \mathcal{M}$. We denote by $\xi$, the local coordinates on $\Sigma_{\tilde{p}+1}$. The maps  $X^M(\xi)$ transform as a vector on the target space and as scalars on the worldvolume. The bosonic M2-brane is obtained by particularizing $\tilde{p}=2$, being $\Sigma_3$ the $(2+1)$ the M2-brane worldvolume and $\mathcal{M}=\mathcal{M}_{11}$ the eleven dimensional Minkowski spacetime. The M2-brane acts as a source for 11D supergravity. it also admits also a Polyakov type action, see \cite{howe1977locally}, \cite{Polyakov:1981rd} for further details. In addition to the embedding maps, an independent auxiliary metric, $g_{ij}$, with determinant $g$ is included in the worldvolume action,  
\begin{equation}\label{ACTION MEMBRANE}
    S_P(X,g_{ij})=-\frac{1}{2}T_{M_2}\int_{\Sigma_3}d^3\xi \bigg( \sqrt{g}g^{ij}\partial_iX^{M}\partial_jX_{M}-\sqrt{g}\Lambda   \bigg).
\end{equation}
The equation of motion obtained from the variation of the auxiliar metric yields an induced metric in terms of the embedding maps.
\begin{equation}\label{induced}
    g_{ij}=\partial_iX^{M}\partial_jX_{M}.
\end{equation}
Hence, both actions coincide for $\Lambda=\tilde{p}-1$. This action can also be coupled to a \textit{Wess-Zumino} term via the coupling to the pullback of the supergravity three-form 
\begin{align}
    T_{M2}\int_{\Sigma_3}\mathbb{P}(C_{[3]}).
\end{align}
In this paper, we focus on studying the effect of the cancellation of the 't Hooft anomaly at the level of the bosonic membrane without the WZ term. This term will be considered in a subsequent work. Therefore, we will not consider here the introduction of this important term any further.
\newline
The symmetries of this action are now, a local symmetry associated with the invariance under general coordinate transformations \cite{Bergshoeff:1987cm}, and the invariance under a global translational symmetry, 
\begin{equation}\label{SYMMETRIA GLOBAL MEMBRANA ANAISADA}
X^{M}\to X^{M}+\epsilon^{M}.
\end{equation}
This is the symmetry that we will be interested in this study.
By considering the Polyakov type action, we can formulate it in the language of forms as follows
\begin{equation}
    S_{P}(X,g_{ij})=-\frac{T_{M_2}}{2}\int_{\Sigma_3} dX^{M}\wedge \star dX_{M}-\frac{T_{M_2}}{2}\int_{\Sigma_3}\sqrt{g} \omega_{vol}(\Sigma_3) ,
\end{equation}
where the Hodge dual is defined with respect to the worldvolume $\Sigma_3$ metric, $\star dX=\sqrt{g}\epsilon_{jkl}g^{ji}\partial_iXd\sigma^k\wedge d\sigma^l$ and $\omega_{vol}(\Sigma_3)$ is a volume-form in the worldvolume. In the symmetry analysis we will consider $\Sigma_3$ to be a compact manifold without boundary. Also, solely the first term in the previus equation will be involved in the symmetry analysis, since the second term is independent of the embedding map,
\begin{align}\label{ACCION DE POLYAKOV}
    S(X)=-T_{M2}\int_{\Sigma_3}dX^M \wedge \star dX_M.
\end{align}
The equation of motion arising from the variation of the action with respect to $X^{M}$, together with the \textit{de Rham} identity are:
\begin{align}\label{EQ DE MOVIMIENTO POLYAKOV}
    d\star dX^{M}={}&0, \\
    d(d X^{M})={}&0 \label{EQ DE BIANCHI MEMBRANA}.
\end{align}
The variation of the action with respect to the metric leads to (\ref{induced}).
$\Sigma_3$ can be consider as the boundary of a 4-dimensional manifold $\mathcal{D}$ of arbitrary topology.
Following the standard analysis for \textit{HFS}(p), we reinterpret \eqref{EQ DE MOVIMIENTO POLYAKOV} and \eqref{EQ DE BIANCHI MEMBRANA} as the conservation law of higher currents. For this case, we have the following set of currents associated with the \textit{HFS}$(0)$ symmetries\footnote{To obtain a HFS in String theory a similar transformation was considered in \cite{Chatzistavrakidis:2021dqg}.}
\begin{align}\label{CORRIENTES DE HFS(0) PARA POLYAKOV}
    J_{m}^{M} = \star dX^{M}, \qquad J_{w}^{M} = dX^{M}.
\end{align}
where by the sub-index $m$ and $w$ we denote 'monopole' and 'winding' currents, respectively. The precise meaning will become evident soon.
These $p-$form currents are conserved by construction. Besides, they are invariant under the global symmetry given by a (global) shift in the embedding map for the field $X^{M} \to X^{M}+\epsilon^{M}$. This global symmetry can be seeing as generated by the topological operators built as:
\begin{align}\label{OPERADORES DE M2 BRANA SIN B}
    U_{\alpha}^{M}(\mathcal{N})={}&\exp\big( i\alpha\int_{\mathcal{N}} \star dX^{M}\big),\\
    U_{\beta}^{M}(\mathcal{Q})={}&\exp\big(i\beta \int_{\mathcal{Q}}dX^{M} \big).
\end{align}
They naturally act on the states of the theory that correspond to Wilson operators.
Here, is a 2-submanifold $\mathcal{N}$ with $codim(\mathcal{N})=1$ and a 1-submanifold $\mathcal{Q}$ with $codim(\mathcal{Q})=2$. These operators are topological for the cobordism group \( \Omega_0^{G} \), which requires that the equations of motion \eqref{EQ DE MOVIMIENTO POLYAKOV} and \eqref{EQ DE BIANCHI MEMBRANA} hold not only on the submanifolds \( \mathcal{N} \) and \( \mathcal{Q} \), but also on all their equivalence classes given by \( \Omega_0^{G} \).
It is easy to see that the action of these operators is trivial unless the submanifolds $\mathcal{N}$ and $\mathcal{Q}$ are non-trivial cycles, \textit{i.e}, when the 2- and 1-forms are closed. Therefore, the operators act trivially on the states as the identity. In this sense the global symmetries of M2-brane are
\begin{align}
\mathbb{G}^{(1)}\cong\bigoplus_{i=1}^2U(1)^{(1)}_{w,i}  \oplus U(1)_m^{(0)}.
\end{align}
Where each sector corresponds to the winding and monopole contribution, respectively.

%%%%%%%%%%%%%%%%%%%%%%%%%%%%%%%%%%%%%%%%%%%%%%%%%%%%%%%%%%%%%%%%%%%%%%%%%%%%%%%%%%%%%%%
\subsection{Background field coupling and the free-anomaly action}
In order to analyze the 't Hooft anomaly present in this theory, background fields are introduced. They couple with the currents \eqref{CORRIENTES DE HFS(0) PARA POLYAKOV} in the action. They are $\mathcal{B} \in \Omega^{1}(\Sigma_3)$ and $\Tilde{\mathcal{B}} \in \Omega^{2}(\Sigma_3)$, with $\Sigma_3 = \partial \mathcal{D}$. Following the procedure explained in the preceding sections, the modified term of the M2-brane action with background fields is given by:
\begin{align}
     S_{}(X^M,\mathcal{B}^M,\Tilde{ \mathcal{B}}^M)={}&-T_{M2}\int_{\Sigma_3}(dX-\mathcal{B})^M\wedge \star (dX-\Tilde{\mathcal{B}})_M+ \notag \\ {}& + T_{M2}\int_{\Sigma_3}\Tilde{\mathcal{B}}^M\wedge(dX-\mathcal{B})_M.
\end{align}
In order to gauge the global symmetry, the 1-form class $ [\mathcal{B}^M]$ must transform in terms of an exact form defined with the same 0-form parameter as in the \eqref{SYMMETRIA GLOBAL MEMBRANA ANAISADA}. 
This action is not invariant for the non-trivial class $H^{1}(\Sigma_3,\mathbb{R})\otimes H^{2}(\Sigma_3,\mathbb{R})$ given by:
\begin{align}\label{CLASE DE EQUIVALENCIA X B BTILDE}
   [X^M] {}& \triangleq  X^{M} \to X^{M}+\epsilon^M, \\ \label{CLASE DE EQUIVALENCIA X B0}
   [\mathcal{B}^M] {}&  \triangleq \mathcal{B}^{M} \to
     \mathcal{B}^{M}+d\epsilon^{M},\\ \label{CLASE DE EQUIVALENCIA X B1}
    [\tilde{\mathcal{B}}^M] {}& \triangleq \Tilde{\mathcal{B}}^{M} \to \Tilde{\mathcal{B}}^{M}+d\Lambda_1^{M}.
\end{align}
In the interests of simplifying the notation, it is hereby stated that the index M of the fields is to be omitted from this point onwards.
The introduction of these background fields, put in evidence the 't Hooft anomaly since:
\begin{align}
    \delta S_{}(g,X,\mathcal{B},\Tilde{ \mathcal{B}})=-T_{M2}\int_{\Sigma_3}\mathcal{B}\wedge d\Lambda_1.
\end{align}
If we introduce a counterterm for this anomaly, as follows
\begin{align}
    S_{ct}(\mathcal{B},\tilde{\mathcal{B}})={}& T_{M2}\int_{\Sigma_3}\mathcal{B}\wedge\tilde{\mathcal{B}},
\end{align}
the variation of the action containing  the former counterterm, still contains an anomaly associated with the variation:
\begin{align}
    \delta \big( S_{}(g,X,\mathcal{B},\Tilde{ \mathcal{B}})+S_{ct}(\mathcal{B},\Tilde{ \mathcal{B}}) \big)=-T_{M2}\int_{\Sigma_3}\Tilde{\mathcal{B}}\wedge d\epsilon.
\end{align}
It is evident here that the problem of the 't Hooft anomaly arises because the introduction of counterterms exchanges the anomalous term between \((w) \to (m)\) and \((m) \to (w)\). This is similar to how, in \(U(1)\) gauge theory, counterterms interchange the roles of electric and magnetic anomalies. In the present case, the required anomalous counterterm the anomaly theory is given by an inflow action with a BF structure defined in a four dimensional manifold $\mathcal{D}$ such that $\partial \mathcal{D}=\Sigma_3$:
\begin{align}
   \mathcal{T}_{TQFT} \triangleq S_{inflow}(\mathcal{B},\tilde{\mathcal{B}})=-T_{M2}\int_{\mathcal{D}}\mathcal{B}\wedge d\tilde{\mathcal{B}}.
\end{align}
The final action with the inflow term is:
\begin{align}\label{ACCTION FREE ANOMALY}
     S_{}(X,\mathcal{B},\Tilde{\mathcal{B}}){}&+S_{inflow}(\mathcal{B},\Tilde{\mathcal{B}})= -T_{M2}\int_{\Sigma_3}(dX-\mathcal{B})\wedge \star (dX-\mathcal{B}) \notag \\ {}&+T_{M2}\int_{\Sigma_3}\Tilde{\mathcal{B}}\wedge dX -T_{M2}\int_{\mathcal{D}}\mathcal{B}\wedge d\Tilde{\mathcal{B}}.
\end{align}
It is straightforward to prove that the action is invariant under the equivalence class \eqref{CLASE DE EQUIVALENCIA X B BTILDE}, \eqref{CLASE DE EQUIVALENCIA X B0} and \eqref{CLASE DE EQUIVALENCIA X B1}. In the appendix \ref{APENDICE DEMOSTRACION FREE ANOMLA TERM} is explicitly shown that the inflow term directly cancels the anomaly. It is important to note that the on-shell anomaly-free action reproduces the original Polyakov action using the background field's equation of motion, \eqref{ACCION DE POLYAKOV}.
The equations of motion of the modified action in $\Sigma_3$ are:
\begin{align}
    d\star (dX-\mathcal{B})=0, \quad d( dX)=0.
\end{align}
Observe that the introduction of background fields only affects to the so-called monopole current, leaving invariant the winding one,
\begin{align}\label{CORRIENTES DE HFS(1)}
    J_{w} = dX, \qquad J_{m} = \star (dX-\mathcal{B}).
\end{align}

% %%%%%%%%%%%%%%%%%%%%%%%%%%%%%%%%%%%%%%%%%%%%%%%%%%%%%%%%%%%%%%%%%%%%%%%%
\subsection{\texorpdfstring{Compactifying the spacetime: the $M2$-brane on $\mathcal{M}_9 \times T^2$}{Compactifying the spacetime: the M2-brane on M9 x T2}}
In the following we consider a particular background, which has been used extensively in the past to analyze properties of the M2-brane. It has the virtue that is relatively simple but contains a nontrivial topology that allow the presence of monopoles,
\begin{align}
    \mathcal{M}_{11} \to \mathcal{M}_9 \times T^2.
\end{align}
In this target space, there are important changes with respect to the previous analysis given by the action of the topological operators and the modification of the action with background fields defined on $\Sigma_3$ and $\mathcal{D}$. Since we have a compact sector with supp($T^2$), the unique non-trivial operators are given as follow:
\begin{align}\label{OPERADORES DE LA M2 CON BACKGROUND}
    U_{\alpha}^{(w)}(\mathcal{N})={}&\exp \big( i \alpha \int_{\mathcal{N}} \star (d\Tilde{X}-\mathcal{B}) \big), \\
    U_{\beta}^{(m)}(\mathcal{Q})={}&\exp \big( i \beta \int_{\mathcal{Q}}d\Tilde{X} \big).
\end{align}
These operators are unique over the nontrivial cycles, since they act in the compact sector defined by the embedding maps. The second operator,  $U_{\beta}^{(m)}$, is defined in a submanifold of $\Sigma_3$ containing non-trivial cocycles in the worldvolume. The Hodge-decomposition of the q-form background fields associated to the compact sector denoted by $\mathcal{B}^r, \tilde{\mathcal{B}^r}$ into their exact, harmonic and co-exact pieces, $\Omega^{q}(\Sigma_3) \cong Im(d_{q-1})\oplus \mathcal{H}^{q}(\Sigma_3)\oplus Im(\delta_{q+1})$ is the following:
\begin{align}
        \mathcal{B}^r \cong (\mathcal{B}_e^r, \mathcal{B}_h^r,\mathcal{B}_{co}^r), \quad
        \tilde{\mathcal{B}}^r \cong ( \tilde{\mathcal{B}}_e^r, \tilde{\mathcal{B}}^r_h,\tilde{\mathcal{B}}_{co}^r).
\end{align}
Due to the nature of the sector $Im(\delta_{q+1})$, the integral over any basis of cocycles on the submanifolds is zero. Therefore, the only sectors that contribute to our analysis correspond to the closed sector, in particular to the harmonic sector since the exact one is trivial. Additionally, in the analysis of the background field coupling, there is a split of the fields according to the decomposition of the background, $\mathcal{B}^M =( \mathcal{B}^m, \mathcal{B}^r)$. We will only consider the gauging of the global symmetry acting in $T^2$ 
with the equivalence classes given by $[\mathcal{B}^r]=\{ d\epsilon^r \}$ and $[\Tilde{\mathcal{B}}^r]= \{ d\Lambda_1^r\}$ and a constant global shift in $\mathcal{M}_9$ given by $\epsilon^m_0$  with $d\epsilon^m_0=0$.
Within this class, the BF-sector\footnote{$BF(\{ \Phi_I\}, \{ F^{I} \})-\text{sector}$ is compact notation for all class of BF-theories.} of the action takes the following form 
\begin{align}
    BF([X],[\mathcal{B}],[\Tilde{\mathcal{B}}])-\text{sector}={}&BF(X,\mathcal{B},\Tilde{\mathcal{B}})-\text{sector}+T_{M2}\int_{\Sigma_3}d\epsilon^r \wedge \Tilde{\mathcal{B}}_r.
\end{align}
Therefore, the anomaly-free theory is written as follows,
\begin{align}
    S_{T}(X^r,\mathcal{B}^r,\Tilde{\mathcal{B}}^r)={}&-T_{M2}\int_{\Sigma_3}DX^r\wedge \star DX_r+T_{M2}\int_{\Sigma_3}dX^r\wedge \Tilde{\mathcal{B}}_r-\\ \notag
    {}& -T_{M2}\int_{\mathcal{D}}\mathcal{B}^r\wedge d\Tilde{\mathcal{B}}_r+S_p(X^m),
\end{align}
where \( DX^r \) is a covariant derivative in the worldvolume (WV), defined as \( DX^r \triangleq dX^r - \mathcal{B}^r \), and \( S_p(X^m) \) is the Polyakov action evaluated in the $X^m$ fields.  It can be directly demonstrated that this action is invariant under the gauging in the compact sector, see appendix \ref{APENDICE DEMOSTRACION FREE ANOMLA TERM}. 
Now, without losing the global structure and preserving the cancellation of the 't Hooft anomaly, we may consider  \( \tilde{\mathcal{B}}^r \in Z^{2}(\Sigma_3, \mathbb{R}) \) \cite{Brennan:2022tyl}. The total action \( S_T \) becomes:
\begin{align}
    S_T(X^r,\mathcal{B}^r, \Tilde{\mathcal{B}}^r)={}&-T_{M2}\int_{\Sigma_3}\bigg( DX^r\wedge \star DX_r+d(X^r\Tilde{\mathcal{B}}_r)\bigg)- \notag \\ {}&-T_{M2}\int_{\mathcal{D}}\mathcal{B}^r\wedge d \tilde{ \mathcal{B}}_r + S_p(X^m).
\end{align}
The total derivative term on $\Sigma_3$ can be reinterpreted 
in terms of a bundle $\mathcal{G}_1^{\nabla}$-gerbe\footnote{A gerbe bundle or more general, $\mathcal{G}_n^{\nabla}$-gerbe is a generalization of principal bundle $P[G]$, that contains in the follow sense, all structures of usuals fiber bundles:
\begin{align*}
 P[G]-bundles \cong \mathcal{G}_0^{\nabla}-gerbe \subset \mathcal{G}_1^{\nabla}-gerbe \subset \mathcal{G}_2^{\nabla}-gerbe \subset \dots \mathcal{G}_n^{\nabla}-gerbe.   
\end{align*}
 The Deligne cohomology $H^{n+2}_D(\mathcal{M},\mathbb{Z})$ are the cohomology groups that characterize this fiber of higher rank.} with a connection \cite{Murray:1994db}, \cite{Murray:2010mz}. If we consider $dH_2=d(X^{r}\tilde{\mathcal{B}}_r)$ as a 3-form of curvature in $H^{3}_{D}(\Sigma_3,\mathbb{Z})$\footnote{The \(H^{3}_{D}(\Sigma_3, \mathbb{Z})\) is the \textit{Deligne} cohomology group \cite{murray2000bundle}}, we have the following in  (appendix \ref{APENDICE GERBE}):
\begin{align}
    \int_{\Sigma_3}F_{[3]} \in{}& 2\pi \mathbb{Z},\\
    dF_{[3]}={}&0.
\end{align}
Being $H_2$ a connection for the $F_3$ form and $[F_{[3]}]$ is the class defined by the \textit{Dixmier-Douady} class. 
Consequently, the existence of the $\mathcal{G}_1^\nabla$-gerbe, which characterizes the elimination of the 't Hooft anomaly of the membrane with background fields, induces (for $\mathbb{Z} \neq 0$) a non-trivial flux condition in the theory.
 The extension of this result to the supersymmetric formulation of the M2-brane -which a priori is a reasonable conjecture due to its topological nature - would imply the discreteness of the spectrum of the M2-brane for toroidal compactifications of $\mathcal{M}_{11}$.
\newline
Analogously to the case of Maxwell on compact space-time, the cancellation of the t' Hooft anomaly via the introduction of background fluxes, in the case where compact spaces are present, implies the breaking of the global symmetries to their discrete subgroups
\begin{align}
    \mathbb{G}^{(1)} \cong \bigoplus_{i=1}^2 \mathbb{Z}_{N,(w)}^{i}  \oplus \mathbb{Z}_{M,(m)},
\end{align}
where each sector corresponds to the windings and monopole contributions.

%%%%%%%%%%%%%%%%%%%%%%%%%%%%%%%%%%%%%%%%%%%%%%%%%%%%%%%%%%%%%%%%%%%%%%%%%%%%%%%%%%%%%%%%%%%%%%%%%%
\section{Algebra and charged operator for \textit{HSF} of bosonic membrane}
The algebra of operators is given as follows, which is derived from the Ward identity defined by the path integral:
\begin{align}
    \langle U_{\alpha}^{}(\mathcal{N}) U_{\beta}^{}(\mathcal{Q}) \rangle ={}& \exp \big(2\pi i \alpha \beta \, \text{Link}(\mathcal{N} | \mathcal{Q}) \big) \langle U_{\beta}^{}(\mathcal{Q}) U_{\alpha}^{}(\mathcal{N}) \rangle.
\end{align}
Here, the operator \(\text{Link}(\mathcal{N} | \mathcal{Q})\) is a topological invariant valued in \(\mathbb{Z}\). Since there are no more topological operators in the theory, this is the only non-trivial relationship between the operators.
\newline
The action of these operators is given in the same manner as the calculation of the algebra, via the Ward identity. Analogously, let us consider a charged object under the global symmetry defined on \(\Sigma_3\). This 1-codimensional charged operator, denoted by \(L(\mathcal{K})\), with \(\mathcal{K}\) being a submanifold of codimension 1 and a quantized charge \(q \in \mathbb{Z}\) \cite{Brennan:2022tyl}, can be considered as an object in the representation \(\rho_{G}\). Thus, the operators in \eqref{OPERADORES DE M2 BRANA SIN B} act as follows:
\begin{align}
    \langle U_{\alpha}^{r}(\mathcal{N}) L(\mathcal{K}) \rangle ={}& e^{2\pi i \alpha q \, \text{Link}(\mathcal{N} | \mathcal{K})} \langle L(\mathcal{K}) \rangle.
\end{align}
For the \textit{HFS}$(0)$ generated by \(\tilde{U}_{\beta}(\mathcal{Q})\), the natural charged object over this symmetry is a fundamental representation \(\rho_{G}\) as \(\mathcal{O}_{q_0}(\gamma)\) with \(q_0 \in \mathbb{Z}\). The action of this operator on \(\mathcal{O}_{q_0}(\gamma)\) is given by:
\begin{align}
     \langle U_{\alpha}^{r}(\mathcal{N}) \mathcal{O}_{q_0}(\gamma) \rangle ={}& e^{2\pi i \beta q_0 \, \text{Link}(\mathcal{Q} | \gamma)} \langle \mathcal{O}_{q_0}(\gamma) \rangle.
\end{align}
Particularizing to the M2-brane case, the charged objects on which the topological operators act, correspond to a Wilson surface, given by
\begin{align}
    L(\mathcal{K}) &\triangleq \mathcal{W}(\mathcal{C}_{(2)}) = \exp \big( iq_H \int_{\mathcal{C}_{[2]}} H\big),
\end{align}
where \( H \) is the connection of the 3-form flux defined over the $\mathcal{G}_1^{\nabla}$-gerbe of the base \cite{murray2000bundle}, and \( \mathcal{C}_{[2]} \) is a non-trivial two-cocycle. This is a generalization of the Wilson line concept in ordinary gauge symmetries \cite{Brylinski:1993ab},\cite{Baez:2010ya}.

\section{Discussion and Conclusions}
In this paper we study the relation between Higher Form Symmetries, the anomaly cancellation and the quantization conditions over the bosonic M2-brane. 
We have shown that the consistent definition of \textit{HFS} for the Polyakov action of the bosonic M2-brane requires the introduction of higher dimensional background fields through a BF coupling. The cancellation the t' Hooft anomaly due to the presence of background fields requires to add an inflow term. The topological operators associated to the Higher Form Symmetry have been constructed and its algebra computed. The states of the M2-brane theory are described by Wilson surface operators that characterize the global holonomy of the M2-brane generalizing the Wilson loops in gauge theories at the level of the symmetries 
Their action is only nontrivial over the states associated to the compact sector of the M2-brane. 
\newline
The cancellation of the 't Hooft anomaly implies the existence of a Gerbe structure describing the geometry of the M2-brane.  The existence of the $\mathcal{G}_1^\nabla$-gerbe, induces (for $\mathbb{Z} \neq 0$) a non-
trivial flux condition in the theory. 
Interestingly, the projection of this flux condition over the spatial part of the worldvolume of the M2-brane induces a worldvolume flux condition, which can be interpreted in terms of a nontrivial U(1) gauge bundle with a connection describing monopoles. It has been proved that this worldvolume flux condition, which can have different origins, implies the discreteness of the mass operator spectrum of the supersymmetric theory for toroidal backgrounds \cite{Boulton:2002br} . The supersymmetric formulation of the M2-brane is free of Lorentz anomaly for $D=11$ and hence consistent from this point of view. In the supersymmetric M2-brane, it has been shown that a worldvolume flux condition may appear due to an irreducible wrapping that induces a central charge condition \cite{Martin:1997cb} or also due to the presence of a quantized 3-form whose pullback induces a $F_2$ flux over the M2-brane worldvolume \cite{GarciaDelMoral:2018jye},\cite{GarciadelMoral:2020dfs}. Indeed, it can be seen that for certain flux components there exists a duality between both cases \cite{GarciaDelMoral:2018jye}.
More generally, the  bosonic M2-brane action includes the \textit{Wess-Zumino} (WZ) term, 
\begin{align}
    S_{M2B} + WZ \text{-Term} \supset T_{M2} \int_{\Sigma_3} \mathbb{P}(C_{[3]}),
\end{align}
where \( \mathbb{P}(C_{[3]}) \) is the pullback to \( \Sigma_3 \) of the 3-form from supergravity. This theory, analogous to one coupled to Chern-Simons terms, acquires an \textit{ABJ} anomaly in the membrane's worldvolume \cite{Montero:2017yja}. To cancel this anomaly, it is necessary to construct defect operators \( \mathcal{D}_{\alpha}(\mathcal{N}) \), which in turn requires the introduction of non-trivial TQFT-type actions supported on the defect in \( \Sigma_3 \) \cite{Garcia-Valdecasas:2023mis}, \cite{Choi:2022jqy}, \cite{Karasik:2022kkq}, \cite{Hasan:2024aow} ,\cite{Fernandez-Melgarejo:2024ffg}, \cite{Cvetic:2023plv}. The implications of this mechanism in the quantum formulation of the M2-brane has not been discussed in this paper, we will consider it elsewhere. 
\newline
Finally, we conclude that the cancellation of the 't Hooft anomaly at the level of \textit{HFS} of the bosonic M2-brane allows the existence of a flux quantization condition which is related to the discreteness of the  mass spectrum  of the supersymmetric M2-brane for these typw of backgrounds.

\section*{Acknowledgements}
 FCP is supported by Doctorado nacional (ANID) 2023 Scholarship N$21230379$, Semillero funding $SEM18-02$ from U. Antofagasta, and supported as graduate student in the “Doctorado en Física Mención Física-Matemática” Ph.D. program at the Universidad de Antofagasta. MPGM is grateful to the Physics Department of Sciences Faculty at the University of Antofagasta, Chile, for their kind invitation, where part of this work was done.  MPGM is partially supported by the PID2021-125700NB-C21 MCI Spanish Grant. A. R. and FCP  want to thank to SEM18-02 project of the U. Antofagasta.
\appendix
\setcounter{equation}{0}
\renewcommand{\theequation}{A.\arabic{equation}}
\section{Invariance of action with background fields}\label{APENDICE DEMOSTRACION FREE ANOMLA TERM} 
 The kinetic term in the action \eqref{ACCTION FREE ANOMALY} is invariant, while for the second term we have, 
\begin{align}
 \delta \int_{\Sigma_3}\Tilde{\mathcal{B}}\wedge dX={}&\int_{\Sigma_3}\Tilde{\mathcal{B}}\wedge d\epsilon.
\end{align}
Now, with the inflow term:
\begin{align}
    \delta \big( \int_{\Sigma_3}\Tilde{\mathcal{B}}\wedge dX-\mathcal{T}_{TQFT}\big)={}& \int_{\Sigma_3}\Tilde{\mathcal{B}}\wedge d\epsilon-\int_{\mathcal{D}}d(\Tilde{\mathcal{B}}\wedge d\epsilon)\\  
    ={}&0. \notag
\end{align}
Via Stokes' theorem and \(\partial \mathcal{D} = \Sigma_3\), the second integral is defined on \(\Sigma_3\) and cancels the anomaly term. 
\subsection{Cancellation of the anomaly in the compact sector}
Our proposal for the anomaly action with \( \text{supp}(\mathcal{D}) \) is given by:
\begin{align}
    \mathcal{T}_{TQFT} = \gamma \int_{\mathcal{D}} \mathcal{B} \wedge d\tilde{\mathcal{B}}.
\end{align}
For the non-trivial class of the compact sector given by \( [\mathcal{B}^r] \) and \( [\tilde{\mathcal{B}}^r] \), we have the following:
\begin{align}
    \mathcal{T}_{TQFT}' = -\bigg(T_{M2}\int_{\mathcal{D}} d\epsilon^r \wedge d\tilde{\mathcal{B}}_r \bigg) + \mathcal{T}_{TQFT}.
\end{align}
Thus, in the path integral \(\tilde{\mathcal{Z}}(X, \mathcal{B}, \tilde{\mathcal{B}})\), transforming under the equivalence class of the background field gives:
\begin{align}
    \tilde{\mathcal{Z}}([X^r], [\mathcal{B}^r], [\tilde{\mathcal{B}}^r]) ={}& \mathcal{Z}([X^r], [\mathcal{B}^r], [\tilde{\mathcal{B}}^r]) \times \exp(\mathcal{T}_{TQFT}') \notag \\ 
    = {}& \mathcal{Z}(X^r, \mathcal{B}^r, \tilde{\mathcal{B}}^r) \times \exp\big( T_{M2}\int_{\Sigma_3} d\epsilon^r \wedge \tilde{\mathcal{B}}_r\big) \notag \\ {}&  \times \exp(\mathcal{T}_{TQFT}) \notag \\ \notag
    = {}& \mathcal{Z}(X^r, \mathcal{B}^r, \tilde{\mathcal{B}}^r) \times \exp\big( T_{M2}\int_{\Sigma_3} d\epsilon^r \wedge \tilde{\mathcal{B}}_r\big) \notag \\ {}& \times \exp(\mathcal{T}_{TQFT}) \times  \exp  \big(-T_{M2}\int_{\mathcal{D}} d\epsilon^r \wedge d\tilde{\mathcal{B}} \big) \notag \\
    = {}& \mathcal{Z}(X^r, \mathcal{B}^r, \tilde{\mathcal{B}}^r) \times \exp(\mathcal{T}_{TQFT}) \notag \\
    = {}& \tilde{\mathcal{Z}}(X^r, \mathcal{B}^r, \tilde{\mathcal{B}}^r).
\end{align}
In the fourth line, we perform integration by parts on  $\delta \mathcal{T}_{TQFT} $, which results in the cancellation of the anomaly term in \( \Sigma_3 \). With this argument, we demonstrate that the compact sector of the theory is free of anomalies, provided that a \( TQFT \) term is introduced, supported on a 4-manifold whose boundary coincides with the worldvolume \( \Sigma_3 \) \cite{Garcia-Valdecasas:2023mis}.
\section{Gerbe formulation}\label{APENDICE GERBE} 
 \renewcommand{\theequation}{B.\arabic{equation}} 
For $F_{[3]} = dH$ with $H \triangleq X^r \Tilde{\mathcal{B}}_r$ supported in WV, via the Weil theorem \cite{Brylinski:1993ab}, we can impose that:
\begin{align}
    \int_{C_{[3]}} F_{[3]} ={}& 2\pi \mathbb{Z},\\
    dF_{[3]} ={}& 0.
\end{align}
Therefore, over $\Sigma_3$, there exists a 3-fold structure $\mathbb{Y}^{[3]}$ defined over the overlap of three open of the atlas $\mathcal{A}$.
Now we can prove that follow:
\begin{align}
    F_{[3]}= dH_i, \qquad \text{in supp}(U_i),
\end{align}
and 
\begin{align}
    dH_i - dH_j = 0 \qquad \text{in supp($U_i \cap U_j$)}.
\end{align}
This structure for $H_i$ allows us to define, over the 2-fold intersection $U_i \cap U_j$, a 1-form given by:
\begin{align}
    H_i - H_j = d\Lambda_{ij}.
\end{align}
Additionally, we can define a 3-fold over the triple overlap $U_i \cap U_j \cap U_k \neq \{ \emptyset \}$, such that there exists a decomposition:
\begin{align}
    d(\Lambda_{ij} + \Lambda_{jk} + \Lambda_{ki}) = 0,
\end{align}
therefore, there exists a model and a total derivative:
\begin{align}
    \Lambda_{ij} + \Lambda_{jk} + \Lambda_{ki} = d\Tilde{\Lambda}_{ijk}, \quad \Tilde{\Lambda}_{ijk} \in \text{supp}(U_{i}\cap U_{j} \cap U_{k}).
\end{align}
Choosing an atlas decomposition $\mathcal{A}(\Sigma_3)$, valued in $H_3(\Sigma_3, \mathbb{Z})$, we can integrate $F_{[3]}$ as:
\begin{align}
    \int_{\Sigma_3}F_{[3]}={}& \sum_{U_i \cap U_j}\int_{U_i \cap U_j}\big(H_i-H_j \big)\\
    ={}&\sum_{U_i \cap U_j}\int_{U_i \cap U_j}d\Lambda_{ij}\\
    ={}& \sum_{U_i \cap U_j \cap U_k}\int_{U_i \cap U_j \cap U_k} \big(\Lambda_{ij}+\Lambda_{jk}+\Lambda_{ki} \big)\\
    ={}& \sum_{U_i \cap U_j \cap U_k}\int_{U_i \cap U_j \cap U_k} \tilde{\Lambda}_{ijk}\\
    ={}& \int_{C_{0}}\tilde{\Lambda}_{[C_0]}\\
    ={}&2\pi n \in \mathbb{Z}.
\end{align}
Where $C_0 \in H_3(\Sigma_3, \mathbb{Z})$. With these properties, we can interpret $F_{[3]} \in H^3_{D}(\Sigma_3, \mathbb{Z})$ as a curvature in the \textit{Dixmier-Douady} class \cite{Murray:1994db}, for this reason, a non-trivial global structure is inherited over $\Sigma_3$.
 \bibliographystyle{unsrt}

\begin{thebibliography}{10}

\bibitem{Zhang:2023wai}
Hao~Y. Zhang.
\newblock {\em {Generalized Symmetries in Supergravities and Superconformal Field Theories via String Theory}}.
\newblock PhD thesis, UPenn, Philadelphia, Pennsylvania U, 2023. arXiv:2306.07311v1 [hep-th].

\bibitem{Bhardwaj:2023kri}
Lakshya Bhardwaj, Lea~E. Bottini, Ludovic Fraser-Taliente, Liam Gladden, Dewi S.~W. Gould, Arthur Platschorre, and Hannah Tillim.
\newblock {Lectures on generalized symmetries}.
\newblock {\em Phys. Rept.}, 1051:1--87, 2024.  arXiv:2307.07547v2 [hep-th].

\bibitem{Heckman:2024obe}
Heckman, Jonathan J. and McNamara, Jacob and Montero, Miguel and Sharon, Adar and Vafa, Cumrun and Valenzuela, Irene
\newblock {On the Fate of Stringy Non-Invertible Symmetries}.
\newblock { arXiv:2402.00118[hep-th]}.

\bibitem{Cvetic:2021sxm}
Mirjam Cvetic, Markus Dierigl, Ling Lin, and Hao~Y. Zhang.
\newblock {Higher-form symmetries and their anomalies in M-/F-theory duality}.
\newblock {\em Phys. Rev. D}, 104(12):126019, 2021. arXiv:2106.07654v2 [hep-th].

\bibitem{Albertini:2020mdx}
Federica Albertini, Michele Del~Zotto, I\~naki Garc\'\i{}a~Etxebarria, and Saghar~S. Hosseini.
\newblock {Higher Form Symmetries and M-theory}.
\newblock {\em JHEP}, 12:203, 2020. arXiv:2005.12831v2 [hep-th].

\bibitem{Boulton:2010nd}
Boulton, Lyonell and Garcia del Moral, M. P. and Restuccia, Alvaro
\newblock{Spectral properties in supersymmetric matrix models}
\newblock{\em Nucl. Phys. B}, 856,716--747, 2012. arXiv:1011.4791 [hep-th].

\bibitem{Martin:1997cb}
I.~Martin, A.~Restuccia, and Rafael~S. Torrealba.
\newblock {On the stability of compactified D = 11 supermembranes}.
\newblock {\em Nucl. Phys. B}, 521:117--128, 1998.  arXiv:hep-th/9706090v2[hep-th].

\bibitem{Boulton:2002br}
L.~Boulton, M.~P. Garcia~del Moral, and A.~Restuccia.
\newblock {Discreteness of the spectrum of the compactified D = 11 supermembrane with nontrivial winding}.
\newblock {\em Nucl. Phys. B}, 671:343--358, 2003.  arXiv:hep-th/0211047v4[hep-th].

\bibitem{GarciaDelMoral:2018jye}
M.~P. Garcia Del~Moral, C.~Las~Heras, P.~Leon, J.~M. Pena, and A.~Restuccia.
\newblock {M2-branes on a constant flux background}.
\newblock {\em Phys. Lett. B}, 797:134924, 2019. arXiv:1811.11231v3 [hep-th].

\bibitem{GarciadelMoral:2020dfs}
M.~P. Garcia~del Moral, C.~Las~Heras, P.~Leon, J.~M. Pena, and A.~Restuccia.
\newblock {Fluxes, twisted tori, monodromy and $U(1)$ supermembranes}.
\newblock {\em JHEP}, 09:097, 2020.  arXiv:2005.06397v1 [hep-th].

\bibitem{Bars:1987nr}
Itzhak Bars.
\newblock {First Massive Level and Anomalies in the Supermembrane}.
\newblock {\em Nucl. Phys. B}, 308:462--476, 1988.

\bibitem{Bars:1989ba}
Itzhak Bars.
\newblock {Membrane Symmetries and Anomalies}.
\newblock {\em Nucl. Phys. B}, 343:398--417, 1990.

\bibitem{Duff:1987cs}
M.~J. Duff, T.~Inami, C.~N. Pope, E.~Sezgin, and K.~S. Stelle.
\newblock {Semiclassical Quantization of the Supermembrane}.
\newblock {\em Nucl. Phys. B}, 297:515--538, 1988.

\bibitem{deWit:1988xki}
B.~de~Wit, M.~Luscher, and H.~Nicolai.
\newblock {The Supermembrane Is Unstable}.
\newblock {\em Nucl. Phys. B}, 320:135--159, 1989.

\bibitem{Goto:1971ce}
Tetsuo Goto.
\newblock{Relativistic quantum mechanics of one-dimensional mechanical continuum and subsidiary condition of dual resonance model}
\newblock{ \em Prog. Theor. Phys.}, 46;1560--1569, 1971.

\bibitem{Gaiotto:2014kfa}
Davide Gaiotto, Anton Kapustin, Nathan Seiberg, and Brian Willett.
\newblock {Generalized Global Symmetries}.
\newblock {\em JHEP}, 02:172, 2015. arXiv:1412.5148v2 [hep-th].

\bibitem{Gomes:2023ahz}
Pedro R.~S. Gomes.
\newblock {An introduction to higher-form symmetries}.
\newblock {\em SciPost Phys. Lect. Notes}, 74:1, 2023. arXiv:2303.01817v2 [hep-th].

\bibitem{Schafer-Nameki:2023jdn}
Sakura Schafer-Nameki.
\newblock {ICTP lectures on (non-)invertible generalized symmetries}.
\newblock {\em Phys. Rept.}, 1063:1--55, 2024. arXiv:2305.18296v2 [hep-th].

\bibitem{Garcia-Valdecasas:2023mis}
Eduardo Garc\'\i{}a-Valdecasas.
\newblock {Non-invertible symmetries in supergravity}.
\newblock {\em JHEP}, 04:102, 2023. arXiv:2301.00777v1 [hep-th].
 
\bibitem{Karasik:2022kkq}
Avner Karasik.
\newblock {On anomalies and gauging of U(1) non-invertible symmetries in 4d QED}.
\newblock {\em SciPost Phys.}, 15(1):002, 2023.  arXiv:2211.05802v3 [hep-th].

\bibitem{Choi:2022jqy}
Yichul Choi, Ho~Tat Lam, and Shu-Heng Shao.
\newblock {Noninvertible Global Symmetries in the Standard Model}.
\newblock {\em Phys. Rev. Lett.}, 129(16):161601, 2022. arXiv:2205.05086v2 [hep-th].

\bibitem{Fernandez-Melgarejo:2024ffg}
Jose~J. Fernandez-Melgarejo, Giacomo Giorgi, Diego Marques, and J.~A. Rosabal.
\newblock {On Non Invertible Symmetries in Type IIB Supergravity}.
\newblock 7 2024. arXiv:2407.09402v2 [hep-th].

\bibitem{Armas:2023tyx}
Armas, Jay and Jain, Akash.
\newblock{Approximate higher-form symmetries, topological defects, and dynamical phase transitions}.
\newblock{ \em Phys. Rev. D}, 109(4):045019,2024. 
arXiv:2301.09628 [hep-th].

\bibitem{Hasan:2024aow}
Azeem Hasan, Shani Meynet, and Daniele Migliorati.
\newblock {The $SL_2(\mathbb{R})$ duality and the non-invertible $U(1)$ symmetry of Maxwell theory}.
\newblock 5 2024. arXiv:2405.19218v3 [hep-th].

\bibitem{Nambu70}
Y. Nambu.
\newblock{ Duality and Hydrodynamics}.
\newblock 1970. Lectures at the Copenhagen Conference.

\bibitem{Restuccia:1998yx}
A.~Restuccia and J.~Stephany.
\newblock {BF models, duality and bosonization in higher genus surfaces}.
\newblock {\em Phys. Rev. D}, 61:085010, 2000.  arXiv:hep-th/9805075v2[hep-th].

\bibitem{Hull:2024bcl} C.~Hull, M.~L.~Hutt and U.~Lindstr\"om, ``Generalised symmetries in linear gravity,'' arXiv:2409.00178 [hep-th].

\bibitem{Luo:2023ive}
Ran Luo, Qing-Rui Wang, and Yi-Nan Wang.
\newblock {Lecture notes on generalized symmetries and applications}.
\newblock {\em Phys. Rept.}, 1065:1--43, 2024.

\bibitem{Brylinski:1993ab}
J.~L. Brylinski.
\newblock {\em {Loop spaces, characteristic classes and geometric quantization}}.
\newblock 1993.

\bibitem{howe1977locally}
Paul~S Howe and RW~Tucker.
\newblock A locally supersymmetric and reparametrisation invariant action for a spinning membrane.
\newblock {\em Journal of Physics A: Mathematical and General}, 10(9):L155, 1977.

\bibitem{Marquard:1989rd}
Marquard, U. and Kaiser, R. and Scholl, M.
\newblock Lorentz Algebra and Critical Dimension for the Supermembranemembrane.
\newblock {\em Phys. Lett. B}, 227:234--238, 1989.

\bibitem{Polyakov:1981rd}
Alexander~M. Polyakov.
\newblock {Quantum Geometry of Bosonic Strings}.
\newblock {\em Phys. Lett. B}, 103:207--210, 1981.

\bibitem{Bergshoeff:1987cm}
E.~Bergshoeff, E.~Sezgin, and P.~K. Townsend.
\newblock {Supermembranes and Eleven-Dimensional Supergravity}.
\newblock {\em Phys. Lett. B}, 189:75--78, 1987.

\bibitem{Chatzistavrakidis:2021dqg}
Athanasios Chatzistavrakidis, Georgios Karagiannis, and Arash Ranjbar.
\newblock {Duality, Generalized Global Symmetries and Jet Space Isometries}.
\newblock {\em Universe}, 8(1):10, 2021. arXiv:2112.00441v2 [hep-th].

\bibitem{Brennan:2022tyl}
T.~Daniel Brennan, Clay Cordova, and Thomas~T. Dumitrescu.
\newblock {Line Defect Quantum Numbers \& Anomalies}.
\newblock 6 2022. arXiv:2206.15401v1 [hep-th].

\bibitem{Murray:1994db}
Michael~K. Murray.
\newblock {Bundle gerbes}.
\newblock {\em J. Lond. Math. Soc.}, 54:403--416, 1996. arXiv:dg-ga/9407015v1[hep-th]. 

\bibitem{Murray:2010mz}
Michael Murray and Danny Stevenson.
\newblock {A note on bundle gerbes and infinite-dimensionality}.
\newblock 7 2010.  arXiv:1007.4922v1 [math.DG]. 

\bibitem{murray2000bundle}
Michael~K Murray and Daniel Stevenson.
\newblock Bundle gerbes: stable isomorphism and local theory.
\newblock {\em Journal of the London Mathematical Society}, 62(3):925--937, 2000.  arXiv:math/9908135v1 [math.DG].

\bibitem{Baez:2010ya}
John~C. Baez and John Huerta.
\newblock {An Invitation to Higher Gauge Theory}.
\newblock {\em Gen. Rel. Grav.}, 43:2335--2392, 2011. arXiv:1003.4485v2 [hep-th]. 

\bibitem{Montero:2017yja}
Miguel Montero, Angel~M. Uranga, and Irene Valenzuela.
\newblock {A Chern-Simons Pandemic}.
\newblock {\em JHEP}, 07:123, 2017. arXiv:1702.06147v1 [hep-th].

\bibitem{Cvetic:2023plv}
Cveti\v{c}, Mirjam and Heckman, Jonathan J. and H\"ubner, Max and Torres, Ethan
\newblock{Fluxbranes, generalized symmetries, and Verlinde\textquoteright{}s metastable monopole}
\newblock{\em Phys. Rev. D}, 4:109,2024. arXiv:2305.09665v1 [hep-th].

\end{thebibliography}

\end{document}